\documentclass[12pt]{iopart}
 \ifx\pdfoutput\undefined
   \usepackage[dvips]{graphicx}
   \else
   \usepackage[pdftex]{graphicx}
   \fi
   \usepackage{epstopdf}
\usepackage{iopams}  

\begin{document}
\DeclareGraphicsExtensions{.pdf,.gif,.jpg}

\title{$\mathcal{PT}$-symmetry breaking and maximal chirality in a nonuniform $\mathcal{PT}$ symmetric ring}
\author{Derek D. Scott$^1$, Yogesh N. Joglekar$^2$}
\address{Department of Physics, Indiana University Purdue University Indianapolis (IUPUI), Indianapolis, Indiana 46202, USA}
\ead{$^1$ derscott@iupui.edu, $^2$ yojoglek@iupui.edu}
\begin{abstract}
We study the properties of an $N$-site tight-binding ring with parity and time-reversal ($\mathcal{PT}$) symmetric, Hermitian, site-dependent tunneling and a pair of non-Hermitian, $\mathcal{PT}$-symmetric, loss and gain impurities $\pm i\gamma$. The properties of such lattices with open boundary conditions have been intensely explored over the past two years. We numerically investigate the $\mathcal{PT}$-symmetric phase in a ring with a position-dependent tunneling function $t_\alpha(k)=[k(N-k)]^{\alpha/2}$ that, in an open lattice, leads to a strengthened $\mathcal{PT}$-symmetric phase, and 
study the evolution of the $\mathcal{PT}$-symmetric phase from the open chain to a ring. We show that, generally, periodic boundary conditions weaken the $\mathcal{PT}$-symmetric phase, although for experimentally relevant lattice sizes $N \sim 50$, it remains easily accessible. We show that the chirality, quantified by the (magnitude of the) average transverse momentum of a wave packet, shows a maximum at the $\mathcal{PT}$-symmetric threshold. Our results show that although the wavepacket intensity increases monotonically across the $\mathcal{PT}$-breaking threshold, the average momentum decays monotonically on both sides of the threshold. 
\end{abstract}
%\pacs{00.00, 20.00, 42.10}
%\maketitle
%---------------------------------------------------------------------------------------------------------------------------------------------------------------%
\section{Introduction}
\label{sec:intro} 
One of the founding tenets of quantum theory is that the Hamiltonian of a "closed" system is Hermitian with respect to the standard inner product; this property of the Hamiltonian implies that the energy spectrum of the system is purely real, and that its eigenvectors form a complete orthonormal set~\cite{qm1,qm2,qm3}. The quantum theory, based on this and other axioms, has been phenomenally successful in explaining experimental results, and predicting various non-trivial and counterintuitive phenomena. Traditionally, non-Hermitian Hamiltonians, with eigenvalues that have an imaginary part, have been used to model open, dissipative quantum systems such as a resistor~\cite{mbt}. Since the seminal work of Bender and co-workers in the late nineties, over the past decade, a new class of non-Hermitian Hamiltonians has been extensively explored~\cite{r1,r2,r3}. Although these Hamiltonians are not Hermitian under the standard inner product, they are invariant under the combined parity ($\mathcal{P}$) and time-reversal ($\mathcal{T}$) operation, and thus are called $\mathcal{PT}$-symmetric Hamiltonians. The eigenvalues $\epsilon_n$ of such a Hamiltonian are {\it real over a range of parameters}, although the corresponding eigenvectors $|n\rangle$ are not orthogonal under the standard inner product. This parameter space where all eigenvalues are real is called the $\mathcal{PT}$-symmetric phase, and the emergence of the first pair of complex eigenvalues that occurs when the Hamiltonian leaves this parameter space is called $\mathcal{PT}$-symmetry breaking.  Since time-reversal is an anti-linear operator, it follows that when the $\mathcal{PT}$ symmetry is broken, the eigenvectors are no longer simultaneous eigenfunctions of the combined $\mathcal{PT}$ operation. 

Over the past three years, $\mathcal{PT}$-symmetric Hamiltonians for tight-binding lattice models have been extensively investigated~\cite{l1,l2,l3,1dchain}. They have included one dimensional chains with non-Hermitian, $\mathcal{PT}$-symmetric tunneling~\cite{ya}, non-Hermitian $\mathcal{PT}$-symmetric pair of on-site impurity potentials~\cite{mark}, two-dimensional lattices with $\mathcal{PT}$-symmetric potential~\cite{segev}, one-dimensional chain of dimers with $\mathcal{PT}$-symmetric impurities on each dimer~\cite{kottos}, and $\mathcal{PT}$-symmetric spin models~\cite{fring} where the effects of boundary conditions are claimed to be negligible. These theoretical investigations have been accompanied by the experimental exploration of consequences of $\mathcal{PT}$ symmetry breaking in coupled optical waveguides~\cite{expt1,expt2,expt25} and electrical circuits~\cite{expt3}. These ongoing studies have hinted at the deep and rich set of phenomena that follow from non-Hermitian Hamiltonians with balanced loss and gain. Although most of these studies have focused on open chains, some recent works have shown that the effect of boundary conditions on the $\mathcal{PT}$-symmetric phase is non-trivial~\cite{ring1,ring2}. 

The current and emerging promising candidates for $\mathcal{PT}$-symmetric systems - coupled optical waveguides and coupled electrical circuits - are presently focused on the realization of $\mathcal{PT}$-symmetric open chains. For electrical circuits, since the spatial circuit geometry can be easily manipulated, introducing inductive coupling between the ``end sites of an open chain" is relatively straightforward~\cite{expt3}. On the other front, two dimensional (2D) lattices of coupled optical waveguides have been extensively experimentally explored~\cite{2d,2d2} in the context of Anderson localization. Thus, creating a one-dimensional system with periodic boundary conditions - a 2D lattice with only boundary waveguides and no interior waveguides - is feasible with current fabrication technology. Therefore, realization and experimental investigation of one dimensional $\mathcal{PT}$-symmetric rings is likely to occur in near future.  

In this paper, we numerically investigate the $\mathcal{PT}$-symmetric phase diagram in a one dimensional system with periodic boundary condition (a $\mathcal{PT}$-symmetric ring). We consider an $N$-site lattice with a position-dependent, parity-symmetric tunneling function $t_\alpha(k)=t_0[k(N-k)]^{\alpha/2}$, and a pair of balanced gain and loss impurities $\pm i\gamma$. In the Hermitian limit, $\gamma=0$, such non-uniform open chain shows tunable energy spectra, wave packet evolution, and Hanbury-Brown-Twiss correlations~\cite{clint}. Our primary results are a follows: i) For $\alpha>0$, we find that the $\mathcal{PT}$-symmetric phase is generally weakened in comparison with its open chain counterpart. However, for experimentally relevant system sizes $N\sim 50$, the critical impurity strength $\gamma_{PT}$ is of the same order of magnitude as its open-chain counterpart. ii) For $\alpha<0$, $\gamma_{PT}$ for a ring is substantially  enhanced from its open chain value for all impurity locations. iii) The average transverse momentum of a wave packet shows a maximum at the $\mathcal{PT}$-symmetry breaking threshold, and decays monotonically on the two sides of it. 

The plan of the paper is as follows. In Sec.~\ref{sec:tb}, we introduce the tight-binding model and discuss the differences between properties of an open chain and a ring that become relevant for small $N$. In Sec.~\ref{sec:phase}, we present the results for the $\mathcal{PT}$-symmetric phase diagram as a function of the scale factor $0\leq\lambda\leq 1$ that determines the tunneling between end points of an  open chain. In Sec.~\ref{sec:chirality}, we discuss the time evolution of a wave packet across the $\mathcal{PT}$-symmetric threshold. We show that the chirality, which encodes the average clockwise or counterclockwise motion of the wave packet around the ring, is quantified by the average momentum and shows a maximum at the threshold. The paper is concluded with a brief discussion and open questions in Sec.~\ref{sec:disc}. Throughout this paper, we consider coupled optical waveguides with complex refractive index and gain as the prototype realization of a $\mathcal{PT}$-symmetric ring~\cite{expt1,expt2,expt25}. 
%---------------------------------------------------------------------------------------------------------------------------------------------------------------%

\section{Tight-Binding Model}
\label{sec:tb}
We start with a one-dimensional, tight-binding, $N$-site lattice with two non-Hermitian impurities $(+i \gamma,-i \gamma )$ located at mirror-symmetric sites $(m,\bar{m})$, where $1\leq m\leq N/2$ and $\bar{m}=N+1-m>m$.  The Hermitian, position-dependent tunneling Hamiltonian for the open chain is 
\begin{equation}
\label{eq:kinetic}
\hat{H}_{0} =-\sum_{i=1}^{N-1} t_{\alpha}(i)(a_{i+1}^{\dagger}a_{i}+a_{i}^{\dagger}a_{i+1}),
\end{equation}
where $t_\alpha(k)=t_0\left[k(N-k)\right]^{\alpha/2}=t_\alpha(N-k)>0$ is the tunneling between sites $k$ and $k+1$, and $a^{\dagger}_k$ ($a_k$) represents the creation (annihilation) operator for a single-particle state $|k\rangle$ localized on site $k=1,\ldots,N$. The difference equation for the coefficients of an eigenfunction $|\psi\rangle=\sum_{k=1}^N f_k |k\rangle$ of the Hamiltonian Eq.(\ref{eq:kinetic}) with eigenenergy $\epsilon$ is given by 
\begin{equation}
\label{eq:diff}
t_\alpha(k)f_{k+1}+t_\alpha(k-1)f_{k-1}=-\epsilon f_k, 
\end{equation}
where open boundary conditions imply $t_\alpha(N)=0=t_\alpha(0)$. We remind the reader that since the tunneling amplitudes  is site dependent, traditional methods for solving Eq.(\ref{eq:diff}) such as the Bethe ansatz are not applicable. Apart from a few exceptions~\cite{l3}, the eigenvalues and eigenvectors of Eq.(\ref{eq:kinetic}) must be obtained numerically. The impurity potential is given by 
\begin{equation}\label{eq:potential}
\hat{V}=i \gamma(a_{m}^{\dagger}a_{m}-a^{\dagger}_{\bar{m}}a_{\bar{m}})\neq \hat{V}^{\dagger}.
\end{equation}  
The action of the parity operator is given by $\mathcal{P} a^{\dagger}_k \mathcal{P} =a^{\dagger}_{\bar{k}}$, where $\bar{k} = N+1-k$ and the anti-linear time-reversal operator 
implies $\mathcal{T} i \mathcal{T}=-i$.  From these definitions, it follows that the non-Hermitian potential $\hat{V}$ is odd under parity and time-reversal individually, and but obeys $\mathcal{PT} \hat{V} \mathcal{PT} = \hat{V}\neq\hat{V}^\dagger$. The $\mathcal{PT}$-symmetry breaking in such a non-uniform open chain has been extensively studied, albeit only numerically due to the absence of analytical methods that are applicable to a system with site-dependent tunneling~\cite{dy}. The Hamiltonian for a $\mathcal{PT}$-symmetric ring is given by 
\begin{equation}
\label{eq:ring}
\hat{H}(\lambda)=\hat{H}_0+\hat{V}+\lambda t_R \left( a^\dagger_1 a_N + a^\dagger_N a_1\right),
\end{equation}
where we choose the tunneling between the end points of the open chain, sites $1$ and $N$, as $t_R=t_\alpha(1)=t_\alpha(N)$ and the scale-factor $0\leq\lambda\leq 1$ allows us to continuously extrapolate from an open chain to a $\mathcal{PT}$-symmetric ring. Before discussing the results for a ring, $\lambda> 0$, we briefly recall the results for an open chain and establish the terminology~\cite{dy}. In the $\mathcal{PT}$-symmetric phase, the eigenvalue spectrum of an $N$-site open chain is non-degenerate and symmetric about zero, and the eigenfunctions for energies $\pm\epsilon$ are related to each other. The bandwidth $\Delta'_\alpha$ of the energy spectrum scales as $\Delta'_\alpha\sim N^\alpha$ for $\alpha\geq 0$ and $\Delta'_\alpha\sim N^{-|\alpha|/2|}$ for $\alpha<0$ because the bandwidth is determined by the largest tunneling amplitude. We use quarter-bandwidth, $\Delta_\alpha\equiv\Delta'_\alpha/4$ as the energy scale and note that when $\alpha=0$, the threshold impurity strength is given by $\gamma_{PT}/\Delta_{\alpha=0}=1$. 

Note that although the distinction between an open chain and a ring is expected to vanish~\cite{fring} in the limit $N\rightarrow\infty$, for small $N$ the differences between the two can be substantial. As an extreme case, let us consider a uniform 3-site lattice with nearest-neighbor tunneling $t_0$. The non-degenerate, particle-hole symmetric spectrum of such an open chain is given by $E=\{-\sqrt{2}t_0,0,+\sqrt{2}t_0\}$. On the other hand, the spectrum of a three-site ring is given by $E=\{-2t_0, t_0,t_0\}$ and is, in general, asymmetric about zero and degenerate. 

\begin{figure}[tbh]
\begin{center}
\begin{minipage}{18cm}
\begin{minipage}{9cm}
\hspace{-1cm}
\includegraphics[angle=0,width=9cm]{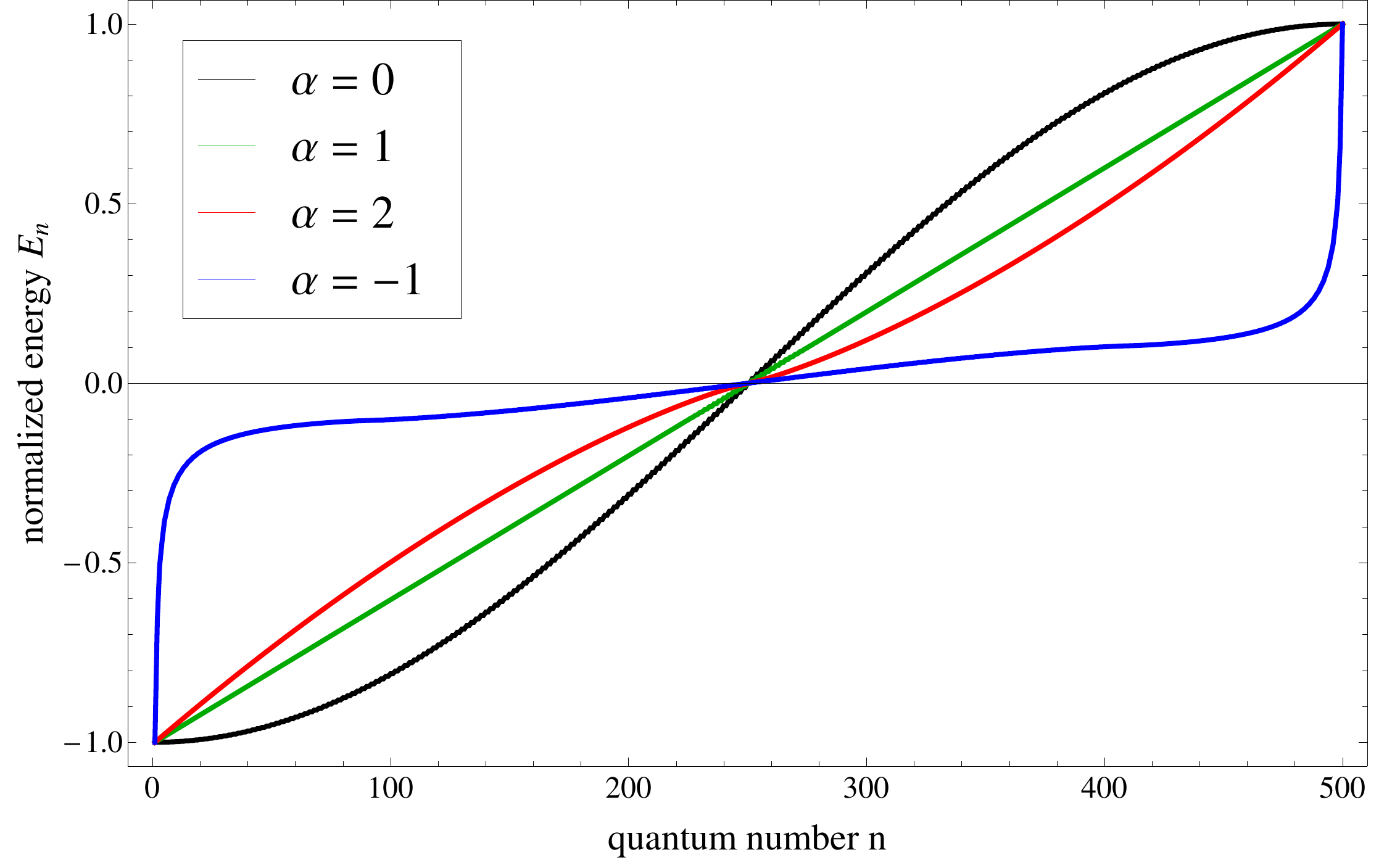}
\end{minipage}
\hspace{-10mm}
\begin{minipage}{9cm}
\includegraphics[angle=0,width=9cm]{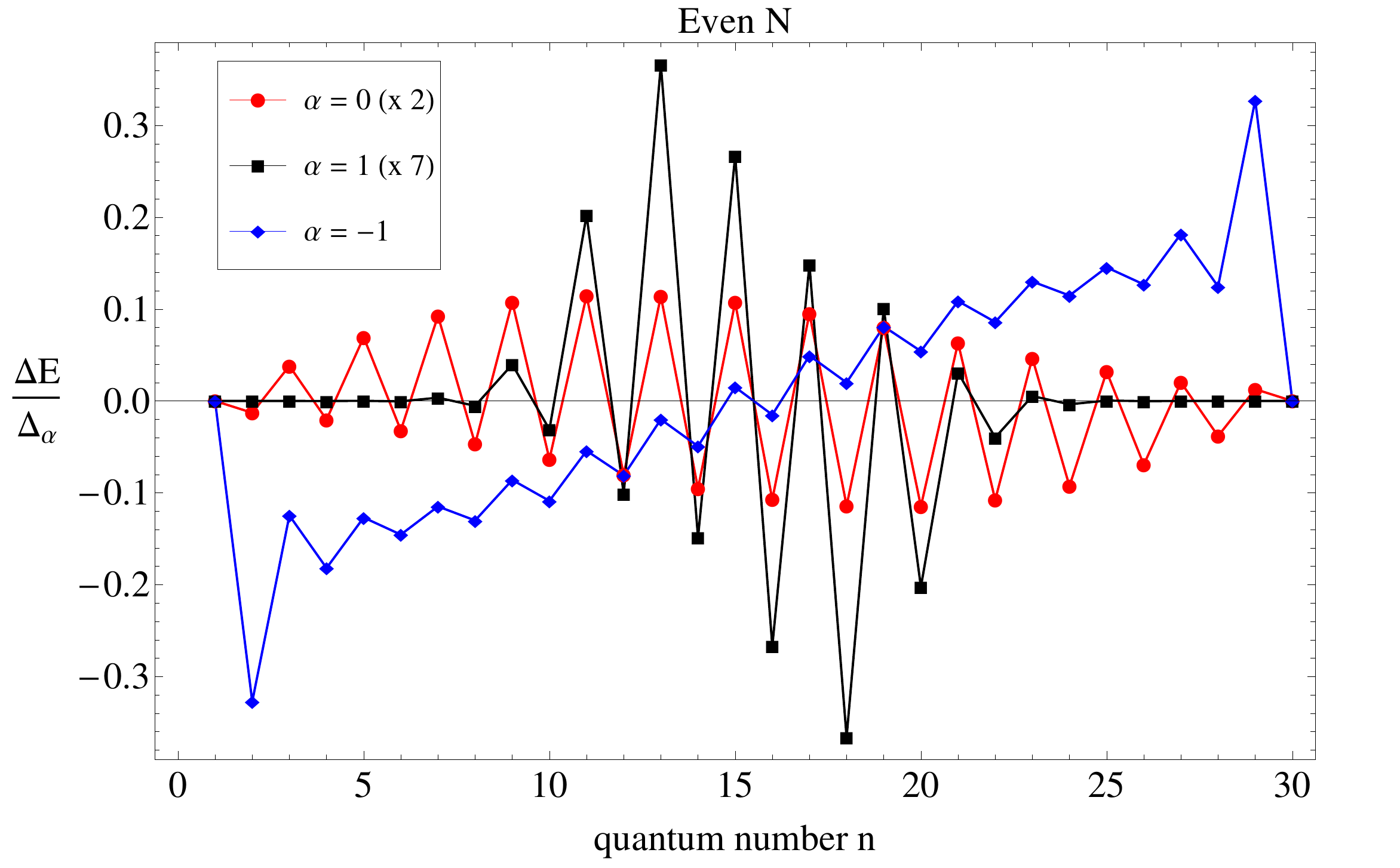}
\end{minipage}
\end{minipage}
\caption{The left-hand panel shows the spectra for Hamiltonian~\ref{eq:kinetic} with $N=500$ and $\alpha=\{-1,0,1,2\}$, where the energy is normalized by its maximum value. The right-hand panel shows the difference between energy spectra for a ring and an open chain for an $N=30$ lattice.  When $\alpha > 0$ the spectral differences are most pronounced near the center of the band. For $\alpha < 0$, the spectral differences are greatest at the band edge and represent the changes that occur in eigenstates localized at the two ends of the open chain~\cite{clint}. These results show that energy spectrum of a $\mathcal{PT}$-symmetric ring is different from that of an open chain for experimentally relevant lattice sizes.}
\label{fig:energy}
\end{center}
\end{figure}
The left-hand panel in Fig.~\ref{fig:energy} shows typical spectra for Hamiltonian~\ref{eq:kinetic} with $N=500$ and $\alpha=\{-1,0,1,2\}$. The energy is normalized by its maximum value and since $N\gg 1$, the spectrum is virtually identical for an open chain and a ring. When $\alpha=0$ (black line) we obtain the expected cosine spectrum, for $\alpha=1$ (green line) and $\alpha=2$ (red line), the spectrum is quasilinear, and when $\alpha=-1$ (blue line), the spectrum band-edges represent eigenstates localized at the end of the chain that are generically present when $\alpha<0$~\cite{clint}. The right-hand panel in Fig.~\ref{fig:energy} shows the difference between energies $\Delta E=E_\mathrm{ring}-E_\mathrm{chain}$ of a ring and an open chain for a lattice with $N=30$ sites and $\alpha =\{0,1,-1 \}$. We find that for {\it $\alpha > 0$ the difference is greatest near the center of the band, whereas for $\alpha < 0$ the spectral difference is greatest at the band edges.} This is expected since the localized edge-states that exist for $\alpha<0$ are most influenced by the introduction of periodic boundary condition. The spectral differences vanish with increasing $N$ but remain pertinent for experimentally relevant lattice sizes~\cite{al0,al1,al2}.

%---------------------------------------------------------------------------------------------------------------------------------------------------------------%

\section{${\mathcal PT}$-symmetric Phase Diagram}
\label{sec:phase}
We now discuss the numerically obtained $\mathcal{PT}$-symmetric phase diagram for the Hamiltonian $\hat{H}(\lambda)$ as a function of increasing loss and gain impurity strength $\gamma$, fractional location of the gain impurity, $\mu=m/N\leq 1/2$, and the dimensionless scale factor $\lambda$ that extrapolates between an open chain ($\lambda=0$) and a ring ($\lambda=1$). The $\mathcal{PT}$-symmetric phase is called robust if the critical impurity strength measured in units of the energy scale, $\gamma_{PT}/\Delta_\alpha(N)$ is nonzero as $N \rightarrow \infty$; 
it is called fragile if $\gamma_{PT}/\Delta_\alpha(N) \rightarrow 0$ as $N \rightarrow \infty$. In an open chain $\mu=1/N$ corresponds to farthest impurities whereas the in a ring, farthest impurities correspond to $\mu\sim 1/4$. We remind the reader that in an open chain, when $\alpha>0$ the critical value of impurity strength $\gamma_{PT}(\mu)$ decreases as the distance between the impurities increases whereas for $\alpha<0$, the $\mathcal{PT}$-symmetric phase is vanishingly small for almost all values of impurity location $\mu$~\cite{dy}. 

\begin{figure}
\begin{center}
\begin{minipage}{18cm}
\begin{minipage}{9cm}
\hspace{-10mm}
\includegraphics[angle=0,width=9cm]{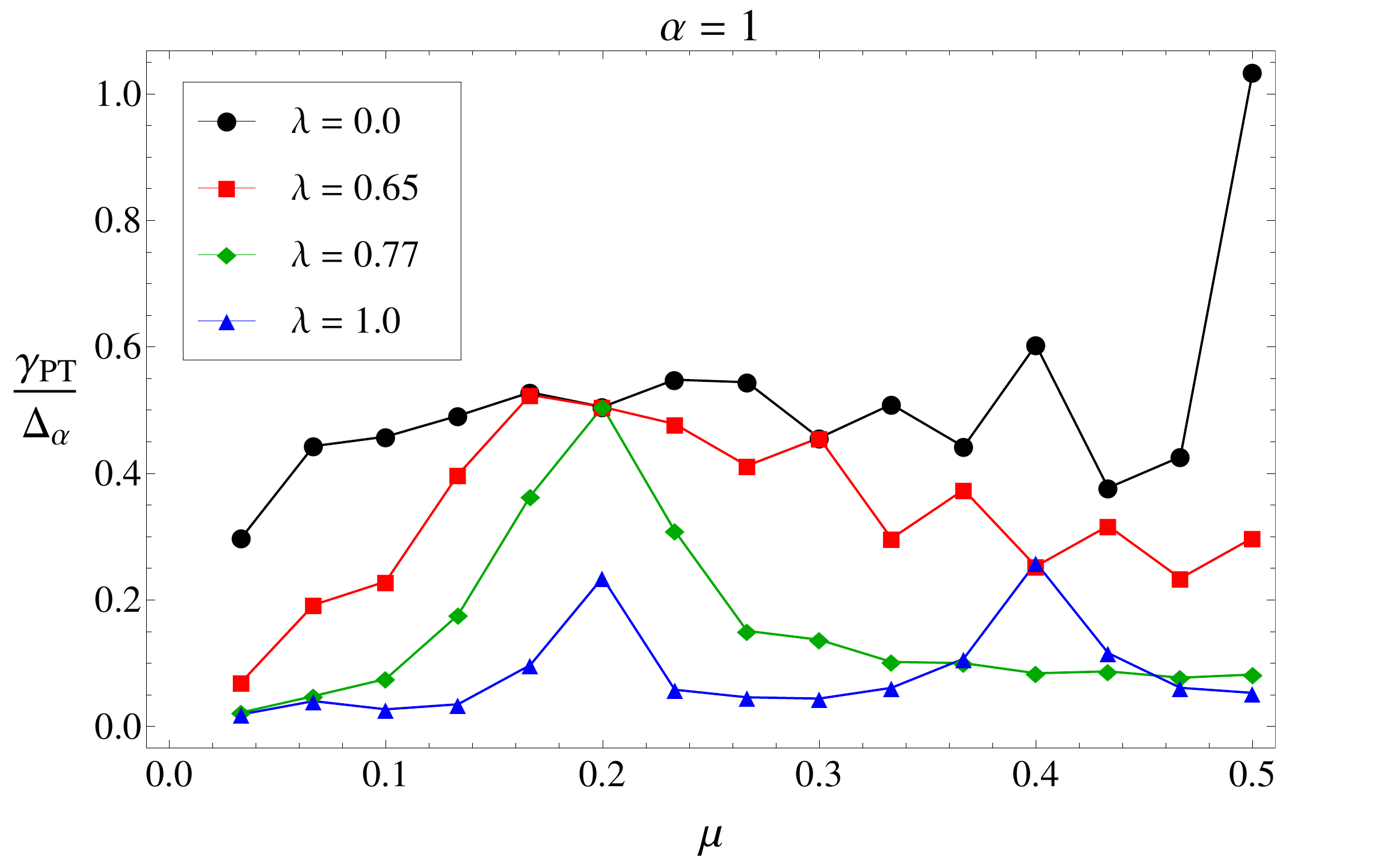}
\end{minipage}
\hspace{-15mm}
\begin{minipage}{9cm}
\includegraphics[angle=0,width=9cm]{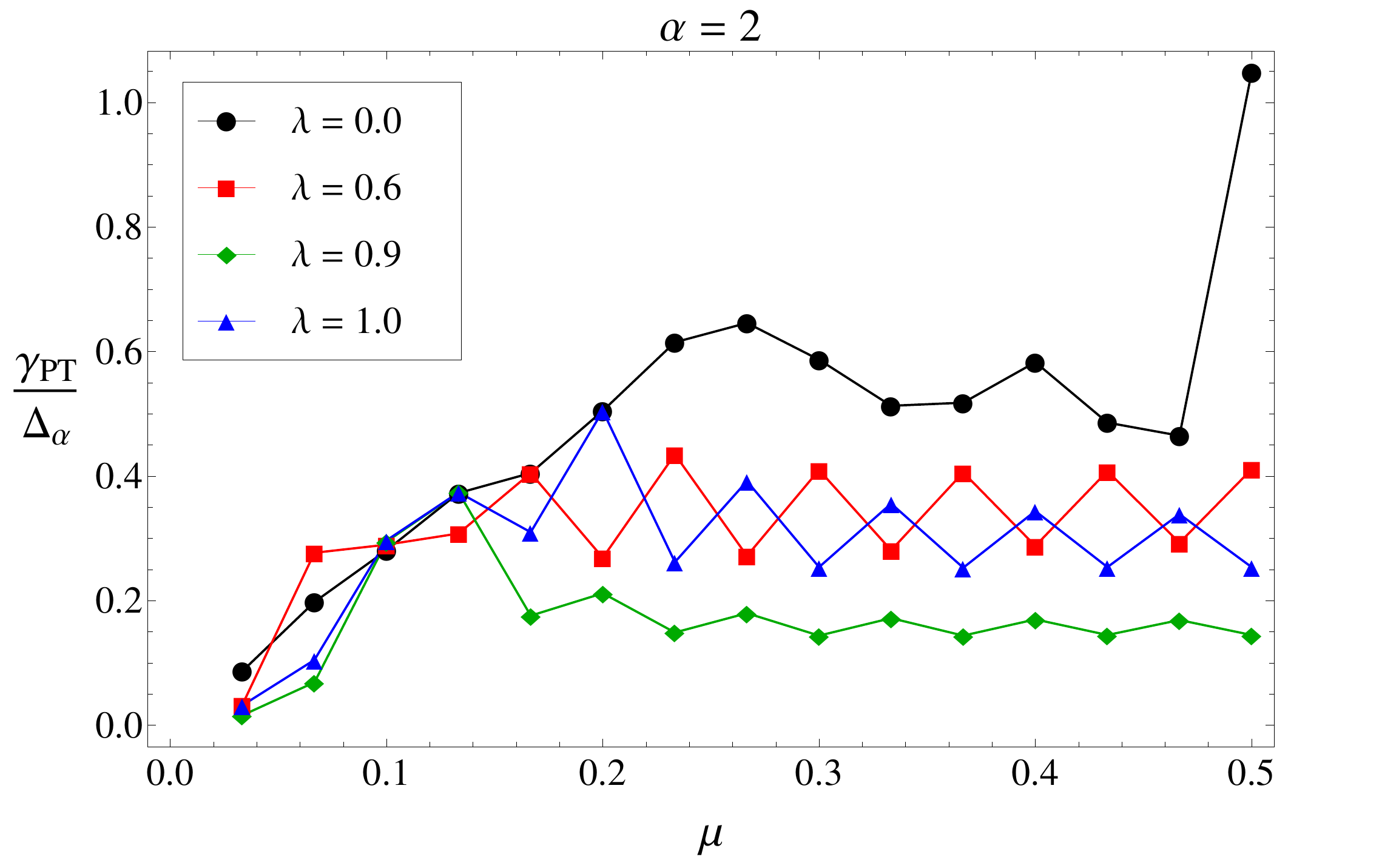}
\end{minipage}
\end{minipage}
\caption{The left-hand panel shows the $\mathcal{PT}$-symmetric phase diagram as a function of impurity strength $\gamma/ \Delta_{\alpha}$ and fractional impurity position $\mu=m/N$ for an $N=30$ site lattice with $\alpha=1$; for $\gamma>\gamma_{PT}$, the eigenvalues of the non-Hermitian Hamiltonian Eq.(\ref{eq:ring}) become complex. The right-hand panel shows corresponding results for the $\alpha=2$ lattice. In both cases, the $\mathcal{PT}$-symmetric phase is maximally robust at $\mu=1/2$ for an open chain, $\lambda = 0$ (black circles). As $\lambda$ is increased, thus increasing the tunneling between sites $1$ and $N$, the critical impurity strength $\gamma_{PT}(\mu)$ remains essentially unchanged from its open chain value for $\lambda\leq 0.5$. As $\lambda$ is increased further (red squares, green diamonds), the $\mathcal{PT}$-symmetric phase in the ring ($\lambda=1$, blue triangles) is weakened for all impurity positions. Thus, a minor change in the Hamiltonian, Eq.(\ref{eq:ring}) leads to a suppression of the critical impurity strength $\gamma_{PT}(\mu)$ even when the impurity location is far away from this change.}
\label{fig:phasediagram}
\end{center}
\end{figure}
Figure~\ref{fig:phasediagram} shows the typical evolution of the $\mathcal{PT}$-symmetric phase diagram in the $(\gamma/\Delta_\alpha,\mu)$ plane as a function of the scale parameter $\lambda$.  These results are for a lattice with $N=30$ sites, and $\alpha=1$ (left-hand panel) and $\alpha=2$ (right-hand panel). For $\lambda=0$ (black circles), the $\mathcal{PT}$-symmetric phase is robust when the loss and gain impurities are closest to each other, $\mu=1/2$. Then the critical impurity strength is given by $\gamma_{PT}/\Delta_\alpha(N)=1$ when $N$ is even and $\gamma_{PT}/\Delta_\alpha(N)=1/2$ when $N$ is odd~\cite{dy}. For $\lambda>0$, we see that the critical impurity strength $\gamma_{PT}$ is, in general, suppressed relative to its value for an open chain for all values of impurity positions $\mu$. Thus, the $\mathcal{PT}$-symmetric phase in a ring with non-uniform tunneling is weaker than its counterpart in an open chain. However, results in Fig.~\ref{fig:phasediagram} also show that the critical impurity strength $\gamma_{PT}$ is still an appreciable fraction of its value in an open chain. The left-hand panel shows that for a lattice with $\alpha=1$, the critical impurity strength $\gamma_{PT}(\mu)$ is strongly suppressed except for a few specific values of $\mu$; this feature is robust irrespective of $N$ and is related to the exactly linear spectrum of an $\alpha=1$ open chain~\cite{l3,ya}. The right-hand panel shows that for $\alpha=2$ lattice, with a non-linear spectrum, the critical impurity strength reaches a plateau $\gamma_{PT}/\Delta_\alpha\sim 0.3$ for most impurity locations $\mu\geq 0.2$. As $\lambda$ is increased from zero (red squares, green diamonds) to one (blue triangles) the $\mathcal{PT}$-symmetric Hamiltonian, Eq.(\ref{eq:ring}), is perturbed only slightly. The tunneling introduced between sites $1$ and $N$, $\lambda t_\alpha(1)=\lambda t_\alpha(N)$, is the smallest among all links for $\alpha>0$. However, {\it it has a dramatic effect on $\gamma_{PT}$ even when the loss and gain impurities $\pm i\gamma$ are away from this tunneling link}, $\mu\geq 0.1$. 

\begin{figure}
\begin{center}
\includegraphics[angle=0,width=9cm]{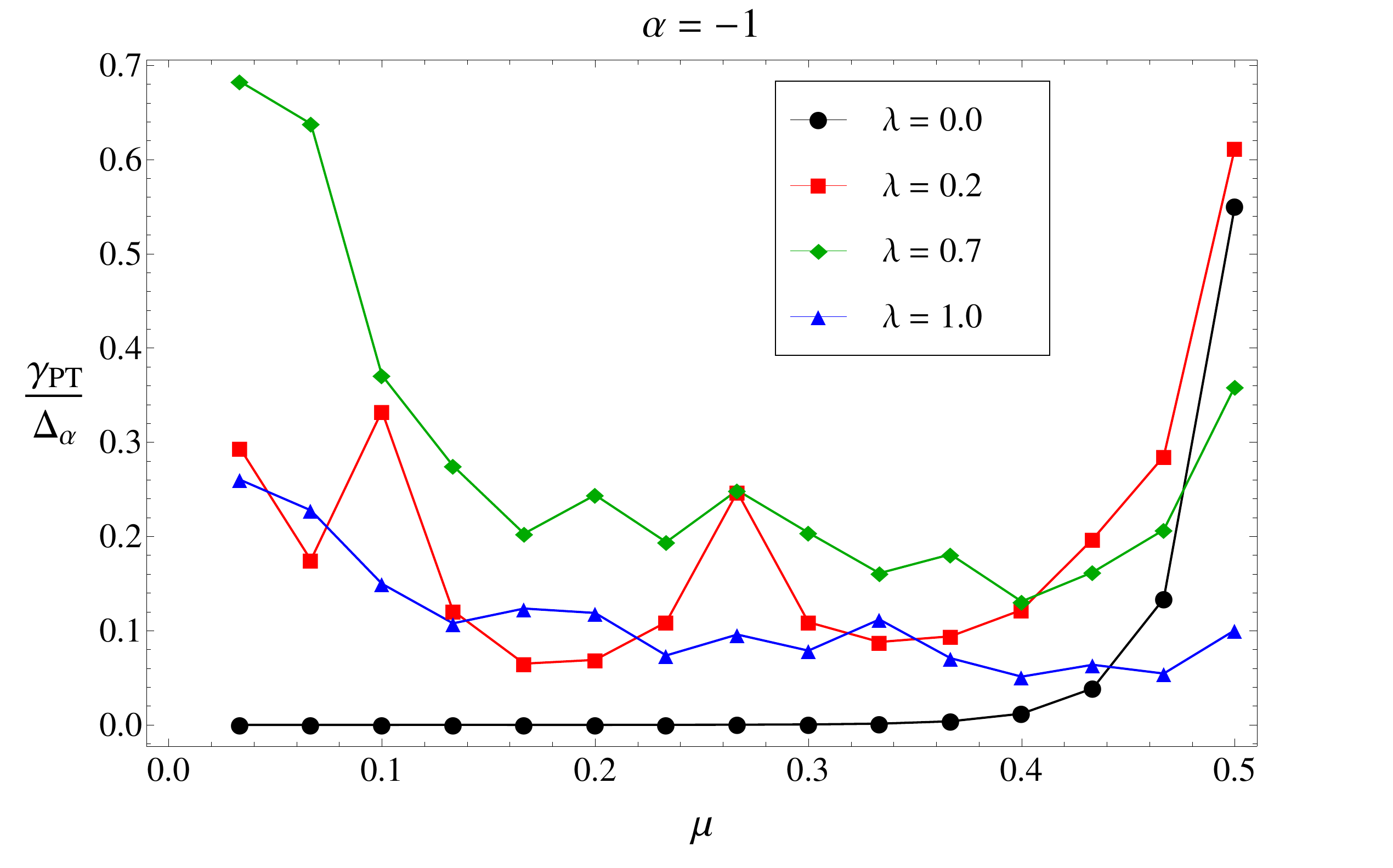}
\caption{$\mathcal{PT}$-symmetric phase diagram for an $N=30$ lattice with $\alpha=-1$, as a function of the scale-factor $\lambda$ that determines the tunneling between sites $1$ and $N$ and extrapolates from an open chain ($\lambda=0$) to a ring ($\lambda=1$). When $\lambda=0$ (black circles), the $\mathcal{PT}$-symmetric phase is fragile everywhere except when the impurities are closest, $\mu=1/2$. For $\lambda>0$, the dimensionless critical impurity strength $\gamma_{PT}/\Delta_\alpha$ is significantly enhanced for all impurity locations $\mu$, and, in contrast with the $\alpha>0$ case, $\gamma_{PT}(\mu)$ for the ring decreases with increasing $\mu$. These results are expected since the tunneling perturbation $\lambda t_\alpha(1)$ scales as the bandwidth for $\alpha<0$ whereas for $\alpha>0$, Fig.~\ref{fig:phasediagram}, the perturbation scales as $\Delta_\alpha/N^{\alpha/2}$.}
\label{fig:phase}
\end{center}
\end{figure}
When $\alpha<0$, the tunneling introduced between sites $1$ and $N$ is the largest among all links and is proportional to the bandwidth $\Delta_\alpha(N)\sim N^{-|\alpha|/2}$. Therefore, we can expect that the fragile $\mathcal{PT}$-symmetric phase in an open chain will be strengthened in a ring. Fig.~\ref{fig:phase} shows the $\mathcal{PT}$-symmetric phase diagram for an $N=30$, $\alpha=-1$ lattice. When $\lambda=0$ (black circles), we obtain the extremely fragile phase diagram of an open chain. As the tunneling is increased from $\lambda=0.2$ (red squares), $\lambda=0.7$ (green diamonds) to $\lambda=1.0$ (blue triangles), we see that the {\it $\mathcal{PT}$-symmetric phase in a ring is substantially strengthened} for all impurity positions. Since $\mathcal{PT}$-symmetric rings ($\lambda=1$) with $\alpha>0$ and $\alpha<0$ can be mapped onto each other with an appropriate redefinition of the impurity location $\mu$, the critical impurity strength $\gamma_{PT}(\mu)$ {\it decreases} with $\mu$ when $\alpha<0$ (Fig.~\ref{fig:phase}) whereas it {\it increases} with $\mu$ for $\alpha>0$ (Fig.~\ref{fig:phasediagram}). 

We remind the reader that in a ring with constant tunneling, accessible to analytical treatment, the critical impurity threshold is {\it zero}~\cite{ring2}. These numerical results show that in a ring with non-uniform tunneling $t_\alpha(i)$ there is a large region, below the blue-triangles-curve in Figs.~\ref{fig:phasediagram} and~\ref{fig:phase}, where the $\mathcal{PT}$-symmetry is exact. 
%---------------------------------------------------------------------------------------------------------------------------------------------------------------%

\section{Chirality across the $\mathcal{PT}$-symmetric Phase Boundary}
\label{sec:chirality}

The time-evolution of a wave packet that is initially localized to a single site has been traditionally used to probe the degrees and signatures of $\mathcal{PT}$-symmetry breaking in coupled optical waveguides~\cite{expt2,expt25}. In these systems, the wave function $|\psi(t)\rangle$ denotes the single-transverse-mode electric field in each waveguide~\cite{al0}, the $\mathcal{PT}$-symmetric impurities which absorb or emit the corresponding electromagnetic radiation are engineered, and the time evolution of an initially normalized wave packet is given by $|\psi(t)\rangle=\exp(-i\hat{H}(\lambda)t/\hbar)|\psi(0)\rangle$ where $\hbar$ is the scaled Planck's constant. The time-evolution operator $\exp(-i\hat{H}(\lambda)t/\hbar)$ is not unitary irrespective of whether the $\mathcal{PT}$-symmetry is exact or broken; therefore, in general, the net intensity $I(t)=\sum_j I(j,t)$ is not one, where $I(j,t)=|\langle j|\psi(t)\rangle| ^2$ denotes the site- and time-dependent intensity. This violation of unitarity is determined by the parity of number of lattice sites $N$~\cite{dy}, and the localized or extended nature of eigenstates~\cite{harsha}. In all cases, however, the net intensity increases monotonically across and exponentially past the $\mathcal{PT}$-symmetric threshold~\cite{kot}. 

The typical time-evolution of an initially localized state shows that apart from spreading across different sites, the wave packet undergoes a preferential clockwise or anti-clockwise motion around the ring. We quantify this tendency, chirality, by a dimensionless, Hermitian, discrete-momentum operator on a ring~\cite{ring2}, 
\begin{equation}
\label{eq:chi}
p_\psi(t)=\langle\psi(t)|\hat{p}|\psi(t)\rangle=-\frac{i}{2}\sum_{j=1}^N\frac{(f^*_{j+1}+f^*_j)(f_{j+1}-f_j)}{|\langle\psi(t)|\psi(t)\rangle|}
\end{equation}
where $|\psi(t)\rangle=\sum_j f_j(t)|j\rangle$ is a time-evolved wave function, and the normalization factor in the denominator is required due to its non-unitary time evolution. Note that due to the Cauchy-Schwartz inequality, the dimensionless momentum satisfies $-1\leq p_\psi(t)\leq 1$. 

When the Hamiltonian is Hermitian and the initial state is localized to a single site, the momentum 
$p_\psi(t)$ symmetrically oscillates about zero. As the impurity strength approaches the threshold value, $\gamma\rightarrow\gamma_{PT}(\mu,\alpha)$, $p_\psi(t)$ remains constant over long time intervals $T\sim 100 N\gg N$ where the time interval $T$ is measured in units of $2\pi\hbar/\Delta_\alpha$~\cite{dy}. We use the time-averaged momentum to denote this steady state value, $p_\psi(\gamma)=\int_{0}^T p_\psi(t') dt'/T$, and choose the time interval $T$ such that the steady-state value $p_\psi(\gamma)$ is independent of it. 

\begin{figure}[htb]
\begin{center}
\begin{minipage}{18cm}
\begin{minipage}{9cm}
%\hspace{-10mm}
%\includegraphics[angle=0,width=9cm]{momentumalpha1.pdf}
\includegraphics[angle=0,width=9cm]{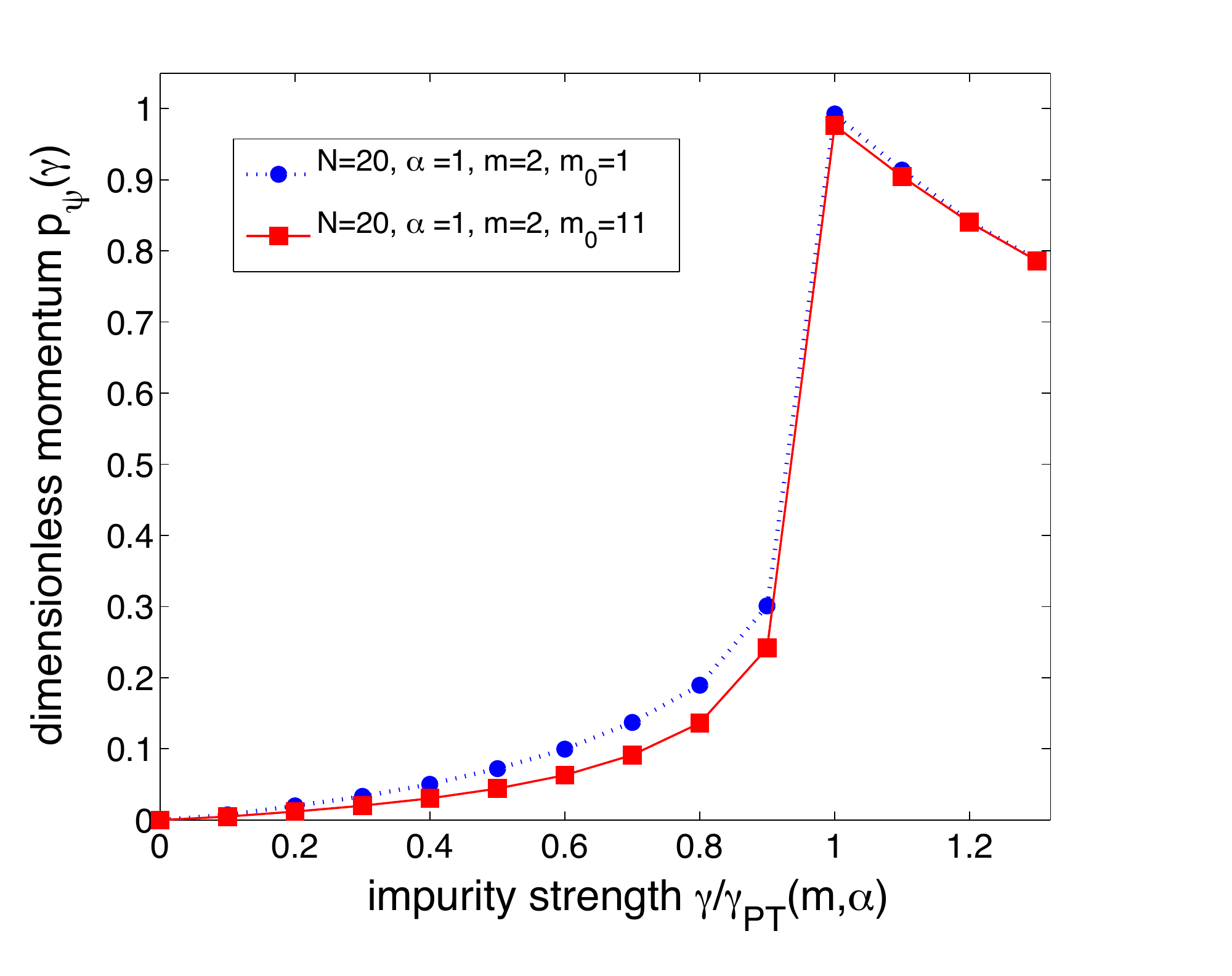}
\end{minipage}
\hspace{-15mm}
\begin{minipage}{9cm}
\includegraphics[angle=0,width=9cm]{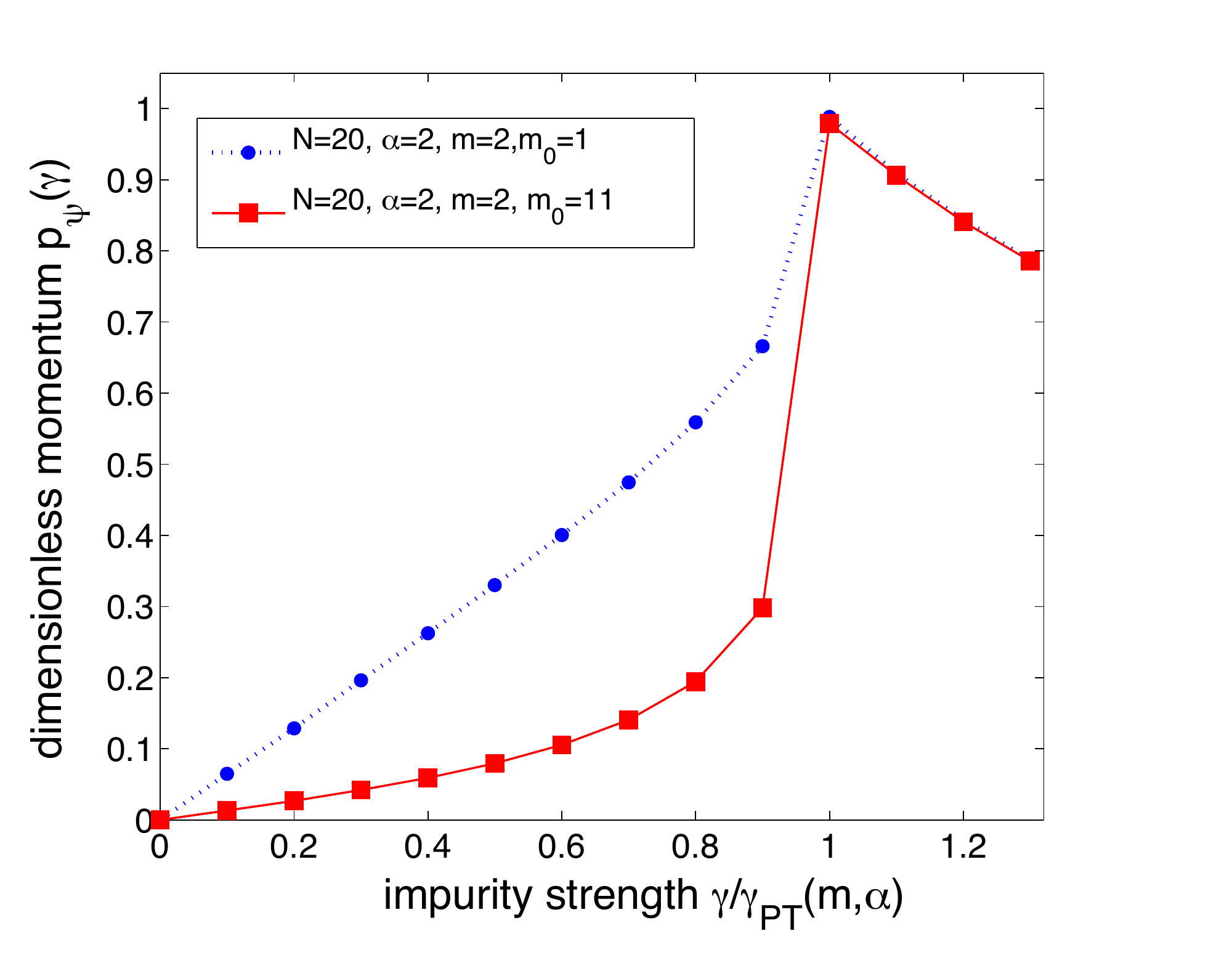}
\end{minipage}
\end{minipage}
\caption{Evolution of the dimensionless, average momentum $p_\psi(\gamma)$ for $N=20$ lattice with gain impurity $+i\gamma$ at position $m=2$ ($\mu=0.1$), and initial wave packet locations $m_0=1$ (blue circles dotted line) and $m_0=11$ (red squares solid line). The left-hand panel corresponds to tunneling profile with $\alpha=1$, whereas the right-hand panel has results for $\alpha=2$. When $\gamma=0$ the average momentum of the wave packet is zero. For small $\gamma$, first-order perturbation theory implies that $p_\psi(\gamma)\propto\gamma$ with a slope that is dependent upon the initial state. In each case, the momentum shows a maximum at the threshold $\gamma=\gamma_{PT}$ and decreases monotonically on both sides of it~\cite{ring2}, although the net intensity increases exponentially beyond the threshold~\cite{kot}.}
\label{fig:maxchiral}
\end{center}
\end{figure}
Figure~\ref{fig:maxchiral} shows the evolution of the dimensionless momentum $p_\psi(\gamma)$ across the $\mathcal{PT}$-symmetric phase boundary for an $N=20$ ring with tunneling functions $\alpha=1$ (left-hand panel) and $\alpha=2$ (right-hand panel). The initial positions of the wave packet are at $m_0=1$ (blue circles dotted line) and $m_0=11$ (red squares solid line). Note that when $\mu=m/N=0.1$, the threshold impurity strengths $\gamma_{PT}(\mu)/\Delta_\alpha$ for $\alpha=1$ and $\alpha=2$ differ by an order of magnitude (Fig~\ref{fig:phasediagram}). In all cases, the average momentum is zero when $\gamma=0$. Its linear increase with $\gamma$ at small $\gamma/\gamma_{PT}(m,\alpha)$ and the different slopes for different initial wave packet locations are both expected from a first-order perturbation theory. Remarkably, in all cases, $p_\psi(\gamma)$ reaches the same maximum possible value (of unity) at the $\mathcal{PT}$-symmetric threshold and decreases monotonically beyond it even though the net intensity $I(t)$ increases exponentially past the threshold. 

At this point, it is worthwhile to recall the corresponding results for a $\mathcal{PT}$-symmetric ring with two (constant) tunnelings between the loss and gain impurities. In that case, the threshold strength $\gamma_{PT}(m)$  is independent of the impurity location $m$, the sign of the momentum $p_\psi(\gamma)$ is determined by the path with the higher tunneling amplitude, and it reaches a universal maximum value of one at the threshold~\cite{dy}. 

\begin{figure}[htb]
\begin{center}
\begin{minipage}{18cm}
\begin{minipage}{9cm}
%\hspace{-10mm}
%\includegraphics[angle=0,width=9cm]{momentumnegative.pdf}
\includegraphics[angle=0,width=9cm]{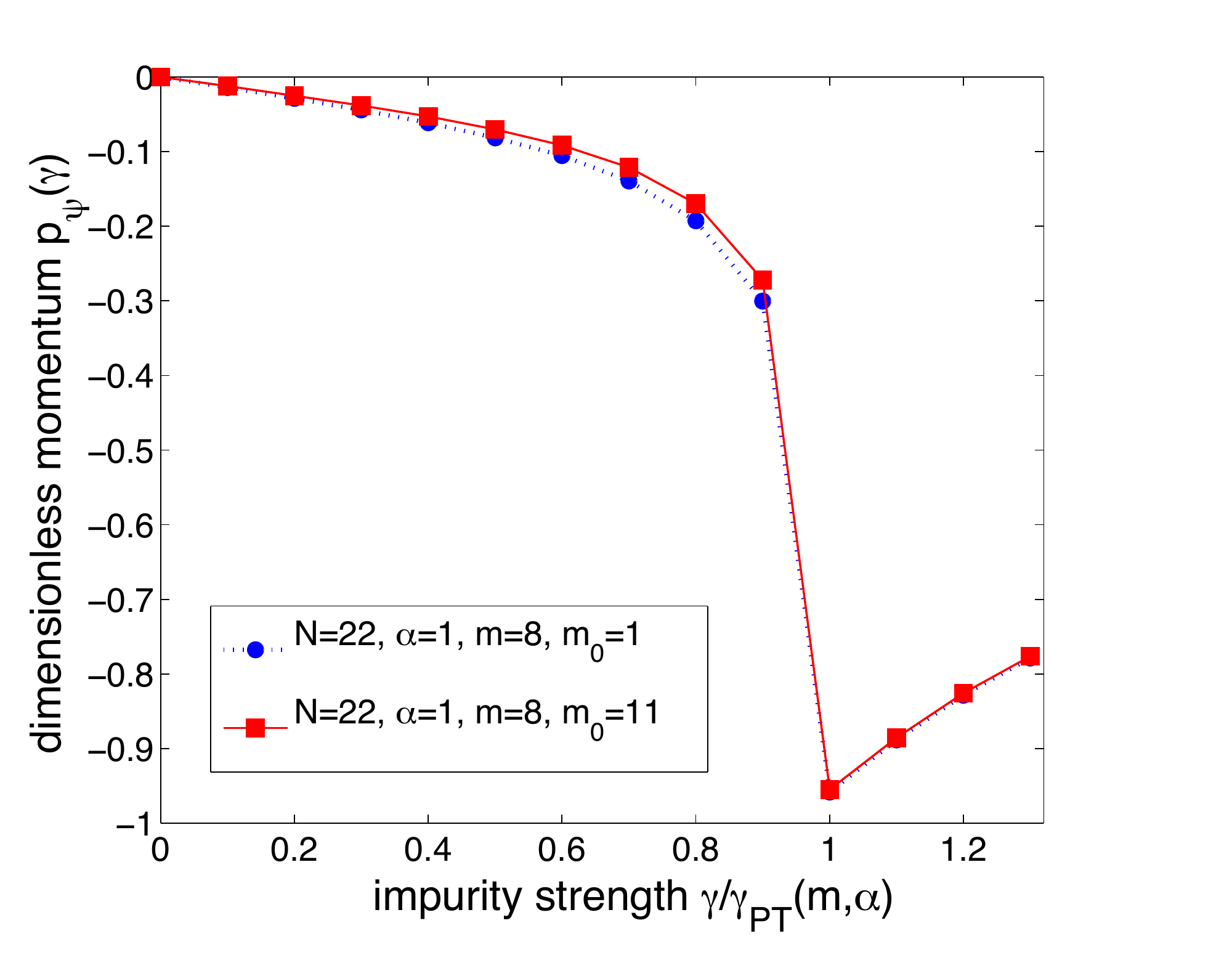}
\end{minipage}
\hspace{-15mm}
\begin{minipage}{9cm}
\includegraphics[angle=0,width=9cm]{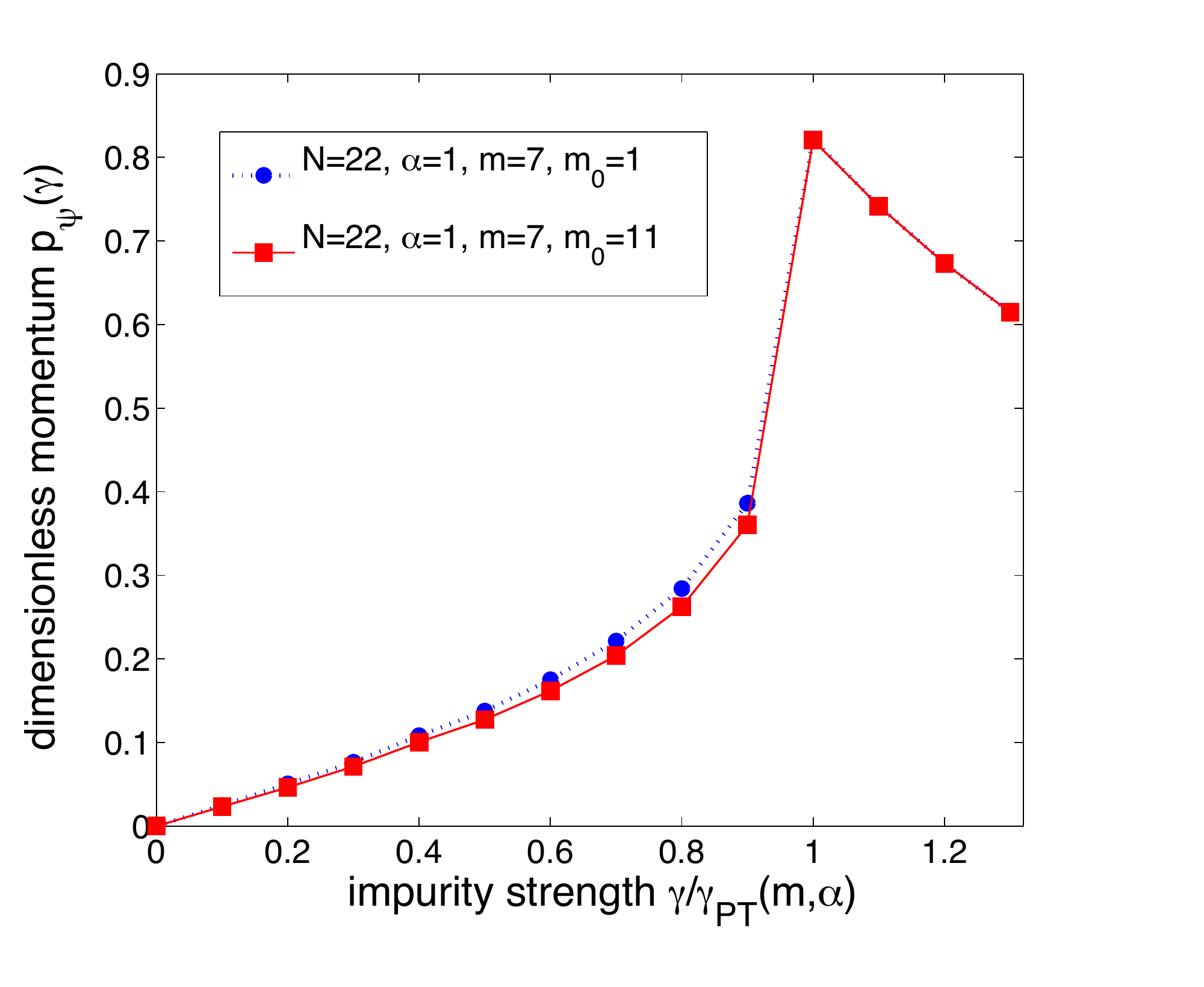}
\end{minipage}
\end{minipage}
\caption{Dimensionless momentum results for an $N=22, \alpha=1$ ring with initial wave packet at locations $m_0=1$ (blue circles dotted line) and $m_0=11$ (red squares solid line). The left-hand panel has gain impurity at $m=8$ ($\mu=0.364$) whereas the right-panel has gain impurity at location $m=7$ ($\mu=0.318$). Due to the position-dependent tunneling Hamiltonian, analytical investigation of momentum dependence on gain impurity location $m$ is not possible. These numerical results show that the sign and the maximum value of the momentum at the threshold are both dependent upon $m$, but not on the initial wave packet location $m_0$.}
\label{fig:nomax}
\end{center}
\end{figure}
Figure~\ref{fig:nomax} shows that for a ring with non-uniform, position-dependent tunneling profile $t_\alpha$, the behavior of the steady-state momentum is not as straightforward. Both panels present  results for an $N=22$ ring with $\alpha=1$ and two initial wave packet locations, $m_0=1$ (blue circles dotted line) and $m_0=11$ (red squares solid line). The left-hand panel has the gain impurity at location $m=8$ ($\mu=0.364$) and it shows that the sign of the steady-state momentum is now negative. The right-hand panel has gain impurity at location $m=7$ ($\mu=0.318$) and it shows that the maximum value attained by the steady-state momentum is not unity. These results imply that, contrary to the two-tunneling ring, sign of the momentum and the maximum value it attains at the threshold are both dependent upon the impurity location. We emphasize, however, that these results are independent of the initial wave packet location, and the {\it qualitative behavior of $p_\psi(\gamma)$ across the $\mathcal{PT}$-symmetric threshold is identical} in all cases. 
%---------------------------------------------------------------------------------------------------------------------------------------------------------------%

\section{Discussion}
\label{sec:disc}
In this paper, we have numerically explored the $\mathcal{PT}$-symmetric phase diagram for a finite, nonuniform $\mathcal{PT}$-symmetric lattice as a function of its boundary conditions. We have shown that for experimentally relevant lattice sizes, the differences between properties of an open chain and a ring are nontrivial, particularly for tunneling profiles $t_\alpha(k)$ with $\alpha<0$. Generically, we found that for $\alpha>0$, the $\mathcal{PT}$-symmetric phase in a ring is weaker than its counterpart in an open chain. In contrast, when $\alpha<0$ the $\mathcal{PT}$-symmetric phase in a ring is substantially strengthened. Since for $\alpha<0$ the tunneling perturbation that is required to change an open chain into a ring is comparable to the bandwidth of the open chain, the strengthening of the $\mathcal{PT}$-symmetric phase is reasonable.

We have shown that the $\mathcal{PT}$ symmetry breaking is accompanied by a qualitatively universal behavior of dimensionless, average momentum $p_\psi(\gamma)$: the momentum is zero when $\gamma=0$, increases linearly with $\gamma$, and its magnitude reaches maximum at the $\mathcal{PT}$-breaking threshold, accompanied by monotonic decay on both sides of the threshold. We have also found that, in contrast with the two-tunneling model~\cite{ring2}, here, the sign of the momentum and its maximum value at the threshold are dependent upon the impurity locations, but not on the initial wave packet location. 

These numerical results raise several questions. For an open chain, the location of the pair of eigenvalues that become degenerate and then complex is uniquely determined by the impurity location $\mu$ irrespective of the value of $\alpha$~\cite{dy}; no such claim seems possible for the $\mathcal{PT}$ symmetry breaking in a ring. The universal presence of the peak in $|p_\psi(\gamma)|$ at the threshold $\gamma=\gamma_{PT}(\mu,\alpha)$ suggests that it may be driven by the exceptional point at the threshold where two eigenvalues become degenerate and the corresponding eigenvectors become parallel~\cite{heiss}. However, lacking analytical methods for the non-uniform tunneling ring, a systematic numerical investigation of the dependence of the sign and the maximum value of the chirality remains an open problem. 
%---------------------------------------------------------------------------------------------------------------------------------------------------------------%

\section{Acknowledgment}
This work was supported by NSF grant DMR-1054020.

%---------------------------------------------------------------------------------------------------------------------------------------------------------------%

%---------------------------------------------------------------------------------------------------------------------------------------------------------------%

\end{document}